\documentclass[prl,twocolumn,amsmath,amssymb]{revtex4}
\usepackage{hyperref}
\usepackage{graphicx}% Include figure files
\begin{document}

\newcommand{\Det}{\,{\rm{Det}\,}}
\newcommand{\Tr}{\,{\rm{Tr}\,}}
\newcommand{\tr}{\,{\rm{tr}\,}}

\newcommand{\bomega}{\bar{\Omega}}
\newcommand{\wsf}{\omega_{\rm sf}}
\newcommand{\sign}{{\rm sign}}

\renewcommand{\Re}{{\rm Re}}
\renewcommand{\Im}{{\rm Im}}

\let\oldmarginpar\marginpar
\renewcommand\marginpar[1]{\-\oldmarginpar[\raggedleft\small\sf #1]{\raggedright\small\sf #1}}

\newcommand{\COMMENT}{\marginpar}
\newcommand{\FigWidth}{\columnwidth}

\title{Anomalous scaling at the quantum critical point}

\author{Ar.~Abanov}
%\email[]{aga@lanl.gov}
%\homepage[]{}
\affiliation{
            Los Alamos National Laboratory,
	    Theoretical division, MS B262,
            Los Alamos, NM 87545
}
\author{A. Chubukov}
\thanks{Present adderss: Dept of Physics, University of Maryland,
College Park, MD, 20742}
\affiliation{
            Department of Physics,
            University of Wisconsin, 
            Madison, WI 53706
}

\date{\today}   
%\date{Last Change \LastChange}% 

\begin{abstract}
We show that  Hertz $\phi^4$
 theory of quantum criticality is incomplete as it misses
 anomalous non-local contributions to the interaction vertices.
 For antiferromagnetic quantum transitions,
 we found that the theory is renormalizable only if the dynamical exponent
  $z=2$. The upper critical dimension is still $d= 4-z =2$, however
 the number of 
 marginal vertices at $d=2$ is infinite. As a result, the theory has
 a finite anomalous exponent already at the upper critical dimension. 
We show that for $d<2$ the Gaussian fixed point splits 
  into two non-Gaussian fixed points.
For both fixed points, the dynamical exponent remains $z=2$.   
\end{abstract}

%\pacs{PACS numbers:75.10.Nr, 05.50.+q, 75.10.Jm}

\maketitle
%\tableofcontents

\begin{figure}
\includegraphics[width=0.6\FigWidth]{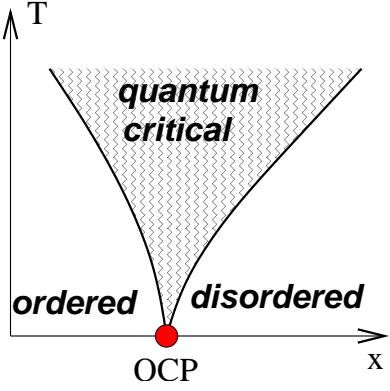}
\caption{\label{fig:hotSpots}
The schematic phase diagram for Quantum Phase Transition.
}
\end{figure}

%%%%%%%%%%%%%%%%%%%%%%%%%%%%%%%%%%%%%%%%%%%%%%%%%%%%%%%%%%%%%%%%%%%
%%%%%%%%%%%%%%%%%%%%%%%%%%%%%%%%%%%%%%%%%%%%%%%%%%%%%%%%%%%%%%%%%%%
%%%%%%%%%%%%%%%%%%%%%%%%%%%%%%%%%%%%%%%%%%%%%%%%%%%%%%%%%%%%%%%%%%%
Quantum phase transitions (QPT) at zero temperature are currently 
 subject
 of intensive experimental and theoretical study. These transitions occur 
 in a number of itinerant fermionic systems 
 under the change of pressure, doping, magnetic field, 
or some other external parameter. QPT 
are very different from 
finite temperature
 phase transitions as the dynamics of the order parameter field can be 
neglected at  finite $T$, but cannot be neglected at $T=0$. 
 The conventional phase
diagram in the $(x,T)$ plane, where $x$ is the external parameter, 
has three distinctive areas (see Fig. \ref{fig:hotSpots}). 
The ordered phase (for most cases, antiferromagnetic) is to the left of the 
 quantum critical point (QCP), 
the disordered, paramagnetic 
 Fermi liquid phase is to the right, and right above 
 QCP there is a quantum critical regime that we will study. 
The system properties in this  regime 
are governed by  quantum dynamics of  slow fluctuations of the 
 order parameter~\cite{sachdev}.

In his original approach to quantum criticality for itinerant fermions, 
  Hertz \cite{hertz} considered coupling between fermions and low-energy 
 bosonic field which condenses at QCP. He
integrated out  fast fermions
 and obtained an effective theory for the slow  bosonic degrees of freedom
 which, he argued, is Landau-Ginsburg-Wilson (LGW) $\phi^4$ theory with  
the upper critical dimension $d_{cr}=4-z$, where $z$ is the 
dynamical exponent. At $d=d_{cr}$, 
the $\phi^{4}$ vertex is marginal, above $d_{cr}$ it is
irrelevant, and for $d < d_{cr}$ it is relevant. Higher order $\phi^6$, etc
 vertices are all irrelevant near $d_{cr}$.
This theory was
latter extended by A. Millis \cite{millis} and others~\cite{others}
to explain the finite temperature properties of metals in the vicinity
of QCP. 

In recent years, the applicability of Hertz-Millis theory to 
 quantum phase transitions in heavy fermion metals has been questioned. 
The theory seems to work in some systems and do not work in others
(see, e.g. \cite{gegenwart}).
It has been suggested that this inconsistency may be due to the fact
 that QPT in some heavy fermion materials require two-band description and 
 may be accompanied by the discontinuous change of the area
of the Fermi surface~\cite{piers}. 
This phenomenon is not included in  Hertz theory
 which describes one-band itinerant models. 

In this paper we consider QPT for which one-band description is valid, and 
 the Fermi surface changes  continuously
 through the transition. We show that even for these transitions, 
 Hertz's theory is incomplete. Specifically, we
 argue that  upon integrating out fermions, the
 effective bosonic theory becomes non-local. 
We show that only the theory with $z=2$ is renormalizable. 
For $d>d_{cr} =2$ all  interaction terms are
irrelevant, and the fixed point is just Gaussian. However, 
 for  $z=2$ and $d=2$,
 {\it there is an infinite number of marginal terms in the effective action}.
 For $d<2$, the Gaussian fixed point splits into 
two new fixed points, one stable and one unstable.
 The scaling dimensions for different components of the momentum are different for these new fixed points, and for both of them,
  the  dynamical exponent $z$ measured in units of the
most relevant component of the momentum equals $2$.

 The existence of  an infinite number of marginal terms at $d=2$
 is in variance with the LGW theory in which only $\phi^4$ vertex is marginal. 
A single marginal vertex gives rise to only 
 logarithmic corrections to a Gaussian theory and does not change  critical 
exponents. This allows one to expand critical exponents in $\epsilon = 2-d$.
 An infinite number of marginal vertices, however, gives rise
 to much stronger, power-law corrections in which case $\epsilon$ 
expansion does not work, and anomalous exponents
 emerge already at the upper critical dimension. This, in particular, 
explains why explicit computations of the dynamical spin susceptibility 
\cite{advances} at the
 antiferromagnetic QCP in $d=2$ yield $\chi (\Omega , {\bf q}) \propto 
(({\bf Q}-{\bf q})^2 + |\Omega_m|)^{-1+\gamma}$ with the 
anomalous exponent $\gamma \approx 0.25$.

Our point of departure is the same as in Hertz theory -- we consider
 fermions coupled to a bosonic field $\phi $ whose 
dynamics is governed by the dynamical exponent $z$. 
The Lagrangian density of this model has a form
\begin{eqnarray}\label{eq:lagrangian}
L&=&\bar{c}_{\omega ,{\bf k}}G_0^{-1}(\omega ,{\bf k})c_{\omega,{\bf k}}+
\frac{1}{2}~\chi_0^{-1}(\Omega ,{\bf q}) \phi^{2}_{\Omega ,{\bf q}} + \nonumber \\
&& g\bar{c}_{\omega ,{\bf k}}c_{\omega -\Omega,{\bf k}-{\bf q}}
\phi_{\Omega ,{\bf q}}
\end{eqnarray}
where $\bar{c}_{\omega ,{\bf k}}$, and $c_{\omega ,{\bf k}}$ 
are Grassmann variables, $g$ is the coupling constant, and
the bare fermionic and bosonic propagators are
\begin{eqnarray}
&&G_0^{-1}(\omega ,{\bf k})=\omega -\epsilon_{\bf k}\label{eq:propG}\\
&&\chi_0^{-1}(\Omega ,{\bf q})=\xi^{-2}+({\bf q}-{\bf Q})^{2}
+(i\Omega)^{2/z}.\label{eq:propChi} 
\end{eqnarray}
Here $\epsilon_{\bf k}$ is the quasiparticle excitation energy, 
$\xi$ is the correlation length which diverges at QCP, 
and ${\bf Q}$ is the ordering vector for 
 $\phi$. An input for the theory is the assumption that there exist 
special hot spots (or lines) at the Fermi surface separated by 
${\bf Q}$.  In the absence of nesting, the Fermi velocities
 at ${\bf k}_{hs}$ and ${\bf k}_{hs}+{\bf Q}$ are not antiparallel.

 The action of Eq. (\ref{eq:lagrangian}) is quadratic in fermions, and 
 Hertz proceeded by integrating  fermions out.
This renormalizes $\chi_0 ( \Omega,{\bf q})$ and also 
introduces the interaction between bosonic fields: $\phi^4, \phi^6$ terms, etc.
Due to particle-hole symmetry, only terms with even number 
of bosonic field are generated.
 The resulting action is \cite{symbolic}
\begin{eqnarray}\label{eq:lagrangianH}
S_{H}=&&\frac{1}{2}
\int d\Omega d^{2}q~
\chi^{-1}(\Omega ,{\bf q})~\phi^{2}_{\Omega ,{\bf q}}+ \nonumber \\
&&\sum_{n=2}^{\infty}\int (d\Omega d^{2}q)^{2n-1}
b_{2n}\left(\phi_{\Omega ,{\bf q}} \right)^{2n} 
\end{eqnarray}

\begin{figure}
\includegraphics[width=\FigWidth]{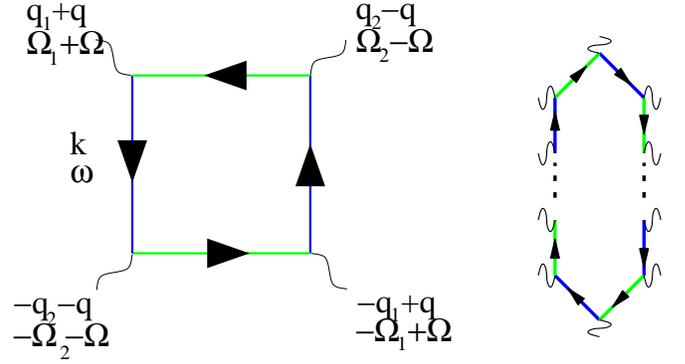}
\caption{\label{fig:Hdiagrams} 
The diagram for the $\phi^{4}$ term in the 
effective bosonic action (left) and  the schematic diagram for
 $\phi^{2n}$ (right). Different colors show Green's
functions near different hot spots.}
\end{figure}

The renormalization of $\chi (\Omega, {\bf q})$ and the vertices $b_{2n}$ 
 are given by the diagrams shown in Fig. \ref{fig:Hdiagrams}. 
Hertz  assumption  that the vertices are local implies 
that they all can be evaluated at bosonic frequencies equal to zero, 
and momenta equal to ${\bf Q}$. 
Consider  $b_{4}$ as an example. 
The corresponding diagram
 contains  two pairs of Green's functions with momenta near 
${\bf k}_{hs}$ and ${\bf k}_{hs} + {\bf Q}$ (see Fig. \ref{fig:Hdiagrams}). 
\begin{equation}\label{eq:b4}
b_{4}\sim 
\int \frac{d\omega d^2k}{(\omega -\epsilon_{\bf k} +i\delta_{\omega })^{2}}
\frac{1}{(\omega - \epsilon_{{\bf k}+{\bf Q}} +i\delta_{\omega } )^{2}}
\end{equation}
One can easily make sure that once we 
 linearize the dispersion near hot spots, the integral over 
momentum would vanish 
 because of double poles~\cite{millis}. 
This implies that the integral in 
(\ref{eq:b4}) comes from 
electrons with high energies, of the order of bandwidth $W$, 
where the spectrum cannot be linearized. Accordingly, the
value of $b_{4}$ scales inversely with $W$ and vanishes in the continuum limit
 $W \rightarrow \infty$.
 The same consideration holds for all other $b_{2n}$.
For finite $W$, all $b_{2n}$ are just some constants. It is then  
 straightforward to 
 calculate the engineering dimensionalities of the
couplings. Using the fact that  $[\Omega ]=z$ (see
 Eq. (\ref{eq:propChi})), we immediately find that in $d=2$ 
  the dimensionality 
of the field is $[\phi^{2}_{\Omega ,{\bf q}}]=-4-z$, and 
 the dimensionalities of the vertices are $[b_{2n}]=2-(n-1)z$.
For $z=2$, this implies that $b_{4}$ is marginal, while
 all the other vertices are irrelevant. This is the known result of 
 the Hertz theory~\cite{hertz,millis}.

We now demonstrate that
 the assumption of the locality of the interaction is in fact incorrect.
Indeed, we found above that the integral in (\ref{eq:b4}) is
zero for a linearized dispersion, because of double poles.
However, this is true only when the bosonic frequencies $\Omega_i$ are zero.
 If we consider instead  the limit $\Omega_i \rightarrow 0$, we  
 find that there exists a tiny range of $|\omega| \leq |\Omega_i|$ where 
 the double poles split into pairs of closely located poles 
 in different half planes. The momentum integration then results in 
small denominators.  
This gives rise to universal, anomalous contributions to 
$b_{2n}$ which may be quite large if the small denominator overshoots the
 smallness of the frequency range.

Linearizing fermionic dispersion near ${\bf k}_{hs}$ and 
${\bf k}_{hs} + {\bf Q}$
 and carrying out integration over momentum and frequency 
in Eq. (\ref{eq:b4}) for  nonzero bosonic 
frequencies $\Omega_i $ and ${\bf q}_i \neq {\bf Q}$, we find 
that 
\begin{equation}
b_4 \propto \frac{g^4}{v^2_F} \frac{|\Omega |}{(\Omega -v_F {\bar q}+i\delta)^{2}}
\label{a_1}
\end{equation}
where ${\bar q} = q -Q$.
The exact expression has a complex dependence on all  
external momenta and frequencies~ \cite{advances}, but
 since we are only interested in the
engineering dimensions, the estimate in (\ref{a_1}) is sufficient.
We see that $b_4$ strongly depends on the ratio $\Omega/v_F q$ and
 actually becomes large in the limit $\Omega\rightarrow 0$, $ v_F q=0$. 
We emphasize that 
 $b_4$ in (\ref{a_1}) is universal in the sense that it does 
not depend on the details of the fermionic dispersion at energies of order $W$
 and survives in the limit $W \rightarrow \infty$.
Restricting with only universal piece in $b_4$, we find that 
$ \phi^4$ term in the effective bosonic action becomes
\begin{equation}\label{eq:g4c}
g^4\int (d^{2}qd\Omega )^{3}\frac{|\Omega |}{(\Omega -v_Fq+i\delta)^{2}}
\left(\phi_{\Omega ,{\bf q}} \right)^{4}
\end{equation}
where $g_{4} \sim g^4/v^2_F$. 
Performing the same calculation for vertices $b_{2n}$ with $n>2$,
we obtain that 
\begin{equation}
b_{2n} \propto \frac{g^{2n}}{v^2_F}~ 
\frac{|\Omega |}{(\Omega -v_Fq+i\delta)^{2(n-1)}}.
\label{a_2}
\end{equation}
Accordingly, $\phi^{2n}$ term in the effective action takes the form
\begin{equation}\label{eq:g4}
g_{2n}\int (d^{2}qd\Omega )^{2n-1}
\frac{|\Omega |}{(\Omega -v_Fq+i\delta)^{2(n-1)}}
\left(\phi_{\Omega ,{\bf q}} \right)^{2n}
\end{equation}
where $g_{2n} \propto g^{2n}/v^2_F$. 

We can now re-evaluate the scaling dimensionality of the vertices.
Performing the same estimates as for the Hertz theory, we obtain
\begin{equation}\label{eq:g4dim}
[g_{2n}]=(2-z)n
\end{equation}
There are two consequences of this result. First, 
Eq. (\ref{eq:g4dim}) holds for all $n$, down to $n=1$.  
For $n=1$, the universal $g_2$ 
 vertex generated by fermions is 
\begin{equation}\label{a_3}
g_{2}\int (d^{2}qd\Omega ) |\Omega |
\left(\phi_{\Omega ,{\bf q}} \right)^{2}
\end{equation}
This is nothing but the Landau damping term. 
Adding the $g_2$ vertex to the Gaussian part of the bosonic action, we 
find that $z=2$ is special in that the form of $\chi_0 (\Omega, {\bf q})$ is  
 reproduced, i.e., the bosonic  dynamics is self-generated. Furthermore, 
 $z=2$ dynamics will obviously dominate even if in the bare
$\chi_0 (\Omega, {\bf q})$ $z <2$. 
Second, we see from  (\ref{eq:g4dim}) that for $z=2$ {\em all}
 vertices are marginal in $d=2$. This
is very different from the Hertz theory where only $\phi^4$  vertex was
marginal. There, a single marginal vertex
lead to only logarithmic corrections to $\chi_0 (\Omega, {\bf q})$ 
 and did not modify critical 
exponents which remain  mean field in $d=2$.
However, when the 
number of marginal vertices is infinite, this argument does not work
  as each vertex
 now gives rise to logarithmic corrections to the susceptibility.
The infinite series of logarithms coming from 
 all $b_{2n}$ sum up into 
 a power-law correction to the susceptibility, such that at $\xi = \infty$,
 \begin{equation}
\chi^{-1} (\Omega, {\bf q}) 
\propto (({\bf q}- {\bf Q})^2 + |\Omega|)^{1-\gamma}
\label{ac_4}
\end{equation}
where $\gamma >0$. This implies that the system develops 
 an anomalous exponent $\gamma$ already at the {\it upper critical dimension}.
 The exponent $\gamma \sim 0.25$ was earlier obtained in perturbative 
$1/N$ calculations  \cite{advances} for Eq. (\ref{eq:propChi}).  
 The present consideration provides understanding of its origin.

\begin{figure}
\includegraphics[width=0.7\FigWidth]{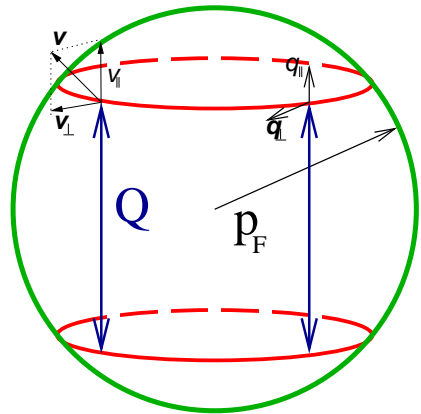}
\caption{\label{fig:manifolds} Hot manifolds in 3D.
}
\end{figure}

We next consider $d \neq 2$.
For Hertz theory,  simple power counting  gives  $[\phi^{2}]=-2-d-z$, and
$[b_{2n}]=(2-d-z)n+d+z$. This implies that
\begin{equation}\label{eq:b4d}
[b_{4}]=4-d-z
\end{equation}
For $z=2$,  $\phi^4$ term  is irrelevant for $d>2$, marginal for $d=2$, and
relevant for $d<2$. All other vertices are  irrelevant near $d =2$.
However, as for $d=2$, this power counting is invalidated because there 
exist  universal, non-analytic contributions to $b_{2n}$ which overshoot 
regular pieces. The calculation of the anomalous pieces in $b_{2n}$
 for $d \neq 2$
 is  a bit more involved compared to $d=2$ case as
in  arbitrary dimension, there exist hot manifolds on
the Fermi surface instead of hot spots  (see Fig. \ref{fig:manifolds}). 
Still, in a single scattering event,  an electron is scattered 
from one hot manifold to another. We therefore can again 
split the Green's functions in the integrals for the vertices
into the two groups belonging to one of two manifolds, and  in each manifold 
 evaluate the anomalous contribution coming from momentum integration 
transverse to the Fermi surface. The integration over the
remaining components of momentum gives the volume of the manifold.
Performing this integration, and also integrating over frequency,
 we obtain 
\begin{equation}
b_{2n} \propto  \frac{|\Omega|} {(\Omega -v_Fq+i\delta )^{2(n-1)}}.
\end{equation}
The $\phi^{2n}$ term in the action then becomes
\begin{equation}\label{eq:g2nd}
g_{2n}\int (d^{d}qd\Omega )^{2n-1}
\left(\phi_{\Omega ,{\bf q}} \right)^{2n}
\end{equation}
Simple power counting now yields 
\begin{equation}\label{eq:d2ndp}
[g_{2n}]=d-2-(d+z-4)n
\end{equation}
We see that  the Gaussian part $[g_{2}]=2-z$ independent of $d$,
i.e.,  $z=2$ dynamics is self generated in any dimension.
We also see that
for $z=2$, $g_{2n}=-(n-1)(d-2)$. This implies that for $d>2$ 
all vertices with $n >1$ are irrelevant, for $d=2$ they are all marginal, 
and for $d<2$ they are all relevant. 
 We see therefore that for $d <2$, the 
 number of relevant vertices is {\it infinite}.
This is is another discrepancy with the Hertz theory.

The case $d<2$ requires further study. 
 Previous analysis of the scaling dimensions of the vertices was pertained to a Gaussian fixed point. Below $d =2$, this fixed point is no longer stable, and we need to find  the new stable fixed point. Our strategy is the following: 
One can straightforwardly verify that for $d<2$, the 
integration over the hot manifold, leading to Eq.
(\ref{eq:g2nd}), affects  differently the  
 components  of the bosonic momenta  parallel to the ordering momentum 
${\bf Q}$ and perpendicular to ${\bf Q}$ ($q_{\parallel}$ and ${\bf q}_{\perp }$, respectively). 
 We therefore introduce an extra scaling dimensionality $\eta$ for one of them.
Using $\eta$ as input, we obtain scaling dimensionalities of the coupling  
constants $g_{2n}$. We then use the fact that at the fixed point 
all vertices must be
 marginal, and obtain the values of $z$ and $\eta$ at the new fixed points.

Let's measure the scaling dimensionalities in units of $q_{\parallel}$.
Then, by definition, $[q_{\parallel}] =1$, while for ${\bf q}_{\perp}$ we
 introduce
\begin{equation}\label{eq:etaPerp}
[{\bf q}_{\perp}]=\eta
\end{equation}
The Gaussian part of the action is 
$\int (d^{d-1}q_{\perp}dq_{\parallel}d\Omega )
(q^{2}_{\parallel}+{\bf q}^{2}_{\perp})\phi^{2}$. 
For $\eta_{\perp }>1$,
 we can neglect ${\bf q}_{\perp}$ in comparison to $q_{\parallel}$,
 while for $\eta_{\perp }<1$ we can neglect $q_{\parallel}$ in comparison to
${\bf q}_{\perp}$. Using this,  we find after a simple algebra 
 that the dimensionality of the field is 
\begin{eqnarray}%
&[\phi^{2}]=-3-(d-1)\eta-z,&\mbox{for $\eta>1$}\nonumber\\%
&[\phi^{2}]=-1-(d+1)\eta-z,&\mbox{for $\eta<1$}\label{eq:phi}
\end{eqnarray}
Using (\ref{eq:phi}), we  then find by straightforward calculations that 
the dimensionalities of the coupling constants
$g_{2n}$ are 
\begin{eqnarray}
[g_{2n}]=&n[3-(d-1)\eta-z]+[(d-1)\eta-1]&\nonumber\\
&\mbox{for $\eta >1$}\nonumber\\
[0pt][g_{2n}]=&n[-1-(d-5)\eta-z]+[(d-3)\eta+1]&\nonumber\\
&\mbox{for $\eta <1$}&\label{eq:g2nde}
\end{eqnarray}
(care has to be taken in regularizing the infrared divergences in the 
 momentum integrals for $\eta <1$). 
At the fixed point
$[g_{2n}]=0$ for all $n$. Solving for $[g_{2n}] =0$, we obtain
\begin{eqnarray}
\left.
\begin{array}{rl}
&\eta=1/(d-1)\\
&z=2
\end{array}
\right|&&\mbox{for $\eta>1$}\label{eq:etaB}\\
\left.
\begin{array}{rl}
&\eta=1/(3-d)\\
&z=2/(3-d)=2\eta
\end{array}
\right|&&\mbox{for $\eta<1$}\label{eq:etaB2}
\end{eqnarray}

We see from (\ref{eq:etaB}, \ref{eq:etaB2}) that each inequality 
is indeed satisfied only if
$d<2$. This implies that for $d<2$ there exist two
fixed points. They progressively deviate from each other as $2-d$ increases.
 Obviously, one of these fixed points is
stable and the other is unstable. To determine which one is stable, we 
 use the result of perturbative $1/N$ analysis of logarithmic
corrections to $v_F q_\parallel$ and $v_F q_\perp$  for $d=2$ ~\cite{advances}.
According to \cite{advances},   the ratio $q_\perp/q_\parallel \propto 1/\log \xi$  renormalizes
  to zero at $\xi = \infty$ (this  makes the Fermi surface nested 
 at the  hot spots at QCP). By continuity, this implies that 
 the stable fixed point is the one with  $\eta>1$.
We see from (\ref{eq:etaB2}) that  at a stable non-Gaussian fixed point
 the dynamical exponent  remains  $z=2$ 

In summary we have demonstrated that Hertz theory of quantum criticality is
incomplete.  We have shown that all vertices 
in the effective bosonic theory possess universal,  anomalous pieces which overshoot constant terms of the Hertz theory.
For quadratic part of the action, this universal contribution is the 
 Landau damping term. We found that $z=2$ is special in that the 
 theory is renormalizable for any $d$. The upper critical dimension for $z=2$
 is $d_{cr} =2$, as in Hertz theory, however,
 the number of marginal vertices at $d=2$ 
is infinite. This gives rise to the appearance of the anomalous 
 exponent already at $d=2$. 
For $d<2$, 
the number of relevant vertices is infinite. 
 We found that the Gaussian fixed point is unstable, and 
there exist two non-Gaussian fixed points  with
 different scaling dimensions for different components of the momentum.
 At each of these new fixed points, there is an infinite number of 
 marginal vertices. We found that one of these points is
 stable and one unstable. Still, for both fixed points, 
the dynamical exponent, measured in units of
 the largest component of the momentum,  remains $z=2$. 

We acknowledge stimulating conversations with A. G. Abanov, 
I. Gruzberg, O. Starykh, and
I. Vekhter. The work of Ar. A. was supported by the Oppenheimer
Fellowship in Los Alamos. A. Ch  acknowledges  
support from NSF DMR 0240238.
We thank the Aspen Center for Physics for hospitality.


\begin{thebibliography}{99}
\bibitem{symbolic} Here and below we use symbolic notations. 
Each field $\phi $ actually depends on  frequency and momentum.
Also, the vertex $b_{2}$ accounts for the
 correction to the correlation 
length. We assume that this correction is already included into 
 $\xi^{-2}$ in Eq. (4).
\end{thebibliography}
\end{document}